\documentclass{llncs}

\usepackage{amsfonts}
\usepackage{color}
\usepackage{url}
\usepackage{ifthen}

\makeatletter
\@ifundefined{lhs2tex.lhs2tex.sty.read}%
  {\@namedef{lhs2tex.lhs2tex.sty.read}{}%
   \newcommand\SkipToFmtEnd{}%
   \newcommand\EndFmtInput{}%
   \long\def\SkipToFmtEnd#1\EndFmtInput{}%
  }\SkipToFmtEnd

\newcommand\ReadOnlyOnce[1]{\@ifundefined{#1}{\@namedef{#1}{}}\SkipToFmtEnd}
\usepackage{amstext}
\usepackage{amssymb}
\usepackage{stmaryrd}
\DeclareFontFamily{OT1}{cmtex}{}
\DeclareFontShape{OT1}{cmtex}{m}{n}
  {<5><6><7><8>cmtex8
   <9>cmtex9
   <10><10.95><12><14.4><17.28><20.74><24.88>cmtex10}{}
\DeclareFontShape{OT1}{cmtex}{m}{it}
  {<-> ssub * cmtt/m/it}{}

\DeclareFontShape{OT1}{cmtt}{bx}{n}
  {<5><6><7><8>cmtt8
   <9>cmbtt9
   <10><10.95><12><14.4><17.28><20.74><24.88>cmbtt10}{}
\DeclareFontShape{OT1}{cmtex}{bx}{n}
  {<-> ssub * cmtt/bx/n}{}
\newcommand{\Conid}[1]{\mathit{#1}}
\newcommand{\Varid}[1]{\mathit{#1}}
\newcommand{\anonymous}{\kern0.06em \vbox{\hrule\@width.5em}}

\usepackage{polytable}

\@ifundefined{mathindent}%
  {\newdimen\mathindent\mathindent\leftmargini}%
  {}%

\def\resethooks{%
  \global\let\SaveRestoreHook\empty
  \global\let\ColumnHook\empty}
\newcommand*{\savecolumns}[1][default]%
  {\g@addto@macro\SaveRestoreHook{\savecolumns[#1]}}
\newcommand*{\restorecolumns}[1][default]%
  {\g@addto@macro\SaveRestoreHook{\restorecolumns[#1]}}
\newcommand*{\aligncolumn}[2]%
  {\g@addto@macro\ColumnHook{\column{#1}{#2}}}

\resethooks

\newcommand{\onelinecommentchars}{\quad-{}- }
\newcommand{\commentbeginchars}{\enskip\{-}
\newcommand{\commentendchars}{-\}\enskip}

\newcommand{\visiblecomments}{%
  \let\onelinecomment=\onelinecommentchars
  \let\commentbegin=\commentbeginchars
  \let\commentend=\commentendchars}

\newcommand{\invisiblecomments}{%
  \let\onelinecomment=\empty
  \let\commentbegin=\empty
  \let\commentend=\empty}

\visiblecomments

\newlength{\blanklineskip}
\newcommand{\hsindent}[1]{\quad}% default is fixed indentation
\let\hspre\empty
\let\hspost\empty

\EndFmtInput
\makeatother
\ReadOnlyOnce{polycode.fmt}%
\makeatletter

\newcommand{\hsnewpar}[1]%
  {{\parskip=0pt\parindent=0pt\par\vskip #1\noindent}}

\newcommand{\hscodestyle}{}

\newcommand{\sethscode}[1]%
  {\expandafter\let\expandafter\hscode\csname #1\endcsname
   \expandafter\let\expandafter\endhscode\csname end#1\endcsname}

  {\par\noindent
   \advance\leftskip\mathindent
   \hscodestyle
   \let\\=\@normalcr
   \let\hspre\(\let\hspost\)%
   \pboxed}%
  {\endpboxed\)%
   \par\noindent
   \ignorespacesafterend}

  {\hsnewpar\abovedisplayskip
   \advance\leftskip\mathindent
   \hscodestyle
   \let\hspre\(\let\hspost\)%
   \pboxed}%
  {\endpboxed%
   \hsnewpar\belowdisplayskip
   \ignorespacesafterend}

  {\hsnewpar\abovedisplayskip
   \advance\leftskip\mathindent
   \hscodestyle
   \let\\=\@normalcr
   \(\pboxed}%
  {\endpboxed\)%
   \hsnewpar\belowdisplayskip
   \ignorespacesafterend}

\newcommand{\plainhs}{\sethscode{plainhscode}}

\plainhs

  {\hsnewpar\abovedisplayskip
   \advance\leftskip\mathindent
   \hscodestyle
   \let\\=\@normalcr
   \(\parray}%
  {\endparray\)%
   \hsnewpar\belowdisplayskip
   \ignorespacesafterend}

  {\parray}{\endparray}

  {\(\parray}{\endparray\)}

\def\codeframewidth{\arrayrulewidth}
\RequirePackage{calc}

  {\parskip=\abovedisplayskip\par\noindent
   \hscodestyle
   \arrayrulewidth=\codeframewidth
   \tabular{@{}|p{\linewidth-2\arraycolsep-2\arrayrulewidth-2pt}|@{}}%
   \hline\framedhslinecorrect\\{-1.5ex}%
   \let\endoflinesave=\\
   \let\\=\@normalcr
   \(\pboxed}%
  {\endpboxed\)%
   \framedhslinecorrect\endoflinesave{.5ex}\hline
   \endtabular
   \parskip=\belowdisplayskip\par\noindent
   \ignorespacesafterend}

\newcommand{\framedhslinecorrect}[2]%
  {#1[#2]}

  {\(\def\column##1##2{}%
   \let\>\undefined\let\<\undefined\let\\\undefined
   \newcommand\>[1][]{}\newcommand\<[1][]{}\newcommand\\[1][]{}%
   \def\fromto##1##2##3{##3}%
   }{\) }%

  {\let\orighscode=\hscode
   \let\origendhscode=\endhscode
   \def\endhscode{\def\hscode{\endgroup\def\@currenvir{hscode}\\}\begingroup}
   \orighscode\def\hscode{\endgroup\def\@currenvir{hscode}}}%
  {\origendhscode
   \global\let\hscode=\orighscode
   \global\let\endhscode=\origendhscode}%

\makeatother
\EndFmtInput

\newboolean{arxiv}
\setboolean{arxiv}{true}

\long\def\symbolfootnote[#1]#2{\begingroup\def\thefootnote{\fnsymbol{footnote}}\footnote[#1]{#2}\endgroup}

\newcommand{\ignore}[1]{}

\newcommand{\activemath}{\mbox{\sc ActiveMath}}
\newcommand{\MathBridge}{Math-Bridge}

\newenvironment{itize}{\begin{list}{--}{\parsep=0pt\parskip=0pt\topsep=1ex\itemsep=0pt}}{\end{list}}

\newcommand{\authorlist}{Bastiaan Heeren${}^{1}$ \and Johan Jeuring${}^{1,2}$}
\newcommand{\address}
    {\begin{tabular}{c}
    ${}^{1}$School of Computer Science, Open Universiteit Nederland \\
    P.O.Box 2960, 6401 DL Heerlen, The Netherlands \\
    \texttt{\{bhr,jje\}@ou.nl}\\
    ${}^{2}$ Department of Information and Computing Sciences, Universiteit Utrecht
    \end{tabular}}
   
\date{}
\title{Adapting Mathematical Domain Reasoners}
\author{\authorlist}\institute{\address}

\begin{document}

\maketitle

\ifthenelse{\boolean{arxiv}}{\symbolfootnote[0]{%
The final publication of this paper is available at \url{www.springerlink.com}}}{}%

\begin{abstract}
Mathematical learning environments help students in mastering mathematical 
knowledge. Mature environments typically offer thousands of interactive 
exercises. Providing feedback to students solving interactive exercises requires 
domain reasoners for doing the exercise-specific calculations. Since a domain 
reasoner has to solve an exercise in the same way a student should solve it, the 
structure of domain reasoners should follow the layered structure of the 
mathematical domains. Furthermore, learners, teachers, and environment builders 
have different requirements for adapting domain reasoners, such as providing 
more details, disallowing or enforcing certain solutions, and combining multiple 
mathematical domains in a new domain. In previous work we have shown how domain 
reasoners for solving interactive exercises can be expressed in terms of rewrite 
strategies, rewrite rules, and views. This paper shows how users can adapt and 
configure such domain reasoners to their own needs. This is achieved by enabling 
users to explicitly communicate the components that are used for solving an 
exercise. 
\end{abstract}

\section{Introduction}
\label{section:intro}

Mathematical learning environments and intelligent tutoring systems such as
MathDox~\cite{mathdox}, the Digital Mathematics Environment (DWO) of the
Freudenthal Institute~\cite{doormanetal2009}, and the {\activemath}
system~\cite{DBLP:conf/icaisc/MelisS04}, help students in mastering mathematical
knowledge. All these systems manage a collection of learning objects, and offer a
wide variety of interactive exercises, together with a graphical user interface
to enter and display mathematical formulas. Sophisticated systems also
have components for exercise generation, for maintaining a student model, for
varying the tutorial strategy, and so on. Mathematical learning environments
often delegate dealing with exercise-specific problems, such as diagnosing
intermediate answers entered by a student and providing feedback, 
to external components. These components can be 
computer algebra systems (CAS) or specialized domain
reasoners.

The wide range of exercise types in a mathematical learning environment 
is challenging for systems that have to
construct a diagnosis from an intermediate student answer to an exercise. In
general, CAS will have no problem calculating an answer to a mathematics
question posed at primary school, high school, or undergraduate university
level. However, CAS are not designed to give detailed diagnoses or suggestions 
to intermediate answers. As a result, giving feedback using CAS is difficult.
Domain reasoners, on the other hand, are designed specifically to give good
feedback.

Developing, offering, and maintaining a collection of domain reasoners for a
mathematical learning environment is more than just a software engineering
problem applied to domain reasoners. Mathematical learning environments usually
offer topics incrementally, building upon prior knowledge. For example, solving
linear equations is treated before and used in solving quadratic equations.
Following Beeson's principles~\cite{mathxpert} of \emph{cognitive fidelity} (the
software solves the problem as a student does) and \emph{glassbox computation}
(you can see how the software solves the problem), domain reasoners should be
organized with the same incremental and layered organization. Structuring domain
reasoners should therefore follow the organization of mathematical knowledge.

Domain reasoners are used by learners, teachers, and developers of mathematical
environments. Users should be able to customize a domain reasoner~\cite{pahl03}.
The different groups of users have various requirements with respect to
customization. For example, a learner might want to see more detail at a
particular point in an exercise, a teacher might want to enforce that an
exercise is solved using a specific approach, and a developer of a
mathematical environment might want to compose a new kind of exercise from
existing parts. Meeting these requirements is challenging in the development of
domain reasoners. It is our experience that users request many customizations,
and it is highly unlikely that a static collection of domain reasoners offering
exercises at a particular level will be sufficient to satisfy everyone.
Instead, we propose a dynamic approach that enables the groups of users to
customize the domain reasoners to their needs.

In this paper we investigate how we can offer users the possibility to adapt and
configure domain reasoners. In the first part of the paper
we identify the problems associated with managing a wide range of
domain reasoners for mathematics, and we argue why allowing configuration 
and adaptation of the concepts describing domain reasoners is desirable. 
This is the paper's first contribution.
Section~\ref{section:motivation} further motivates our research question.
We then give a number of case studies in Section~\ref{section:cases} that 
illustrate the need for adaptation and configuration. Most of these case studies
are taken from our work on developing domain reasoners for about 150 
applets from the DWO of the Freudenthal Institute. 

The second part starts with an overview of the fundamental concepts by means of
which we describe mathematical knowledge for solving exercises in domain
reasoners. We show how these concepts interoperate, and how they are combined
(Section~\ref{section:concepts}). Next, we present a solution for adapting and
configuring domain reasoners in Section~\ref{section:config}, which is our
second contribution. In particular, we show how our solution helps in solving
the case studies. The techniques that are proposed in this paper have been
implemented in our framework for developing domain reasoners\footnote{For more
information, visit our project webpage at \url{http://ideas.cs.uu.nl/}.}, and we
are currently changing the existing domain reasoners accordingly. We evaluate
the advantages and disadvantages of our approach, and draw conclusions in the
final section.

%prevents sloppy paragraph spacing (vertical) on page 2
\ifthenelse{\boolean{arxiv}}{}{\newpage}
 
\section{Motivation}
\label{section:motivation}

Computer algebra systems (CAS) are designed specifically for solving
complex mathematical tasks, and performing symbolic computations. CAS are often
used in intelligent tutoring systems as a back-end for assessing the correctness
of an answer. In general, they are suitable for such a task, although different
normal forms can have subtle effects on an assessment~\cite{1576039}. CAS are
less suitable for supporting more advanced tutoring functionality, such as
suggesting a meaningful next step, showing a worked-out example, or discovering
a common misconception: they have not been designed to do so, and generally
violate the principles of cognitive fidelity and glassbox computation.

Specialized domain reasoners are designed with excellent facilities for feedback
and diagnosis in mind. Because they are specialized they often operate on a
narrow class of exercises (e.g., only linear equations). Supporting more,
related classes (e.g., all mathematics topics covered in high school) raises the
question how the knowledge should be organized and managed. Mathematical
knowledge is typically hierarchical, and according to the principle of cognitive
fidelity, such hierarchies should also be present in a domain reasoner for
mathematics.

\subsection{Feedback services}
\label{services}

When a mathematical learning environment uses domain reasoners for several 
classes of exercises, it is important that the reasoners share a set of
\emph{feedback services}, and that these services are exercise independent. We
have defined such a set of services around rewrite
strategies~\cite{rewritestrategiesforexercises,gerdes-08}, which produce
step-wise solutions for exercises. With a strategy we can
produce worked-out examples (the \emph{derivation} service), suggest a next step
(the \emph{allfirsts} service), and diagnose a term submitted by a learner (the
\emph{diagnose} service). By collecting the rewrite rules of a
strategy, we can report which rules can be applied (the \emph{applicable} service),
or recognize common misconceptions (the \emph{findbuggyrules} service). 
Other services we offer are variations of the ones listed above. All services 
calculate feedback automatically from a strategy specification 
and rewrite rules. 

Goguadze~\cite{DBLP:conf/mkm/Goguadze09} describes a set of feedback services
used by the {\activemath} learning environment to serve as an interface for
calling external domain reasoners. His services are similar to ours, and also
assume the presence of rewrite rules. However, they do not depend on
rewrite strategies. Neither his nor our current services~\cite{gerdes-08} 
accommodate for customizing and adapting domain reasoners.

\subsection{Customization from four perspectives}

Using a predefined collection of domain reasoners that cannot be customized
limits the level of adaptivity of a learning environment. Users of an
environment have many wishes about customizing a domain reasoner, and satisfying
these would lead to many variants. We propose a solution in which users can
adapt a domain reasoner without changing the domain reasoner's implementation.
We identify four perspectives for which we consider customizability and
adaptability. These perspectives correspond to the different groups of users.
\begin{itize}
   \item {\bf Learners.\quad} Learners want to customize an exercise to their own
   level of expertise. They expect guidance at points where
   they experience difficulties. Learners do not interact with a domain 
   reasoner directly, but they send their requests by way of a learning 
   environment.
   \item {\bf Teachers.\quad} Teachers have specific requests about how an
   exercise should be solved, and using which steps. They have a good
   understanding of the capabilities of a particular homogeneous group of
   learners. Teachers want to tailor exercises at a high level.
   \item {\bf Mathematical learning environments.\quad} A learning environment
   is the front-end for practicing mathematical problem solving, and usually
   offers many different classes of exercises. Advanced environments include
   tools for authoring exercises (for teachers), they maintain a model of a
   learner, and can have a component for adaptive course
   generation~\cite{DBLP:conf/ectel/UllrichLM09}. All these aspects are related
   to domain reasoners, and the facilities they offer for customization.
   Environments are the primary clients of a domain reasoner.
   \item {\bf Domain reasoners.\quad} From within a domain reasoner, the main
   concerns are reusability and maintainability of code and components. The
   major issue is how mathematical knowledge should be represented and
   organized, reflecting the layered structure of that knowledge. 
\end{itize}
Each of the case studies that is presented in the next section belongs to one of
the perspectives.
  
\section{Case studies}
\label{section:cases}

This section presents five case studies illustrating the need for dynamic 
domain reasoners that are easily adaptable. Afterwards, we propose a solution,
and revisit the cases in Section~\ref{section:revisit}.

\subsection{Case study: controlling the solutions for an exercise}
\label{case1}

A quadratic equation can be solved in many ways.
For example, the Dutch 
mathematics textbook Getal \& Ruimte~\cite{getalenruimte}, used in more than 
half of the high schools in the Netherlands, gives many techniques to solve an 
equation of the form $ax^2 + bx + c = 0$. It considers the case of a binomial 
($b=0$ or $c=0$) and the case where its factors can be found easily. 
Furthermore, the book shows how \ensuremath{(\Varid{x}\mathbin{+}\mathrm{3})^\mathrm{2}\mathrel{=}\mathrm{16}} can be solved without reworking 
the term on the left-hand side. Of course, the quadratic formula is given as a 
general approach, although using it is discouraged because it is more 
involved.  Figure~\ref{figure:derivations} 
shows alternative derivations for a quadratic equation, including a derivation 
in which the technique of ``completing the square'' is used. Selecting the 
appropriate technique for a given equation is one of the skills that needs 
training.
\begin{figure}[t]
\begin{center}
\begin{minipage}[t]{3cm}
\begin{math}
x^2 - 4x = 12 \\
x^2 - 4x - 12 = 0 \\
(x - 6)(x + 2) = 0 \\
x = 6 \vee x = -2
\end{math}
\end{minipage}\hspace*{3mm}
\begin{minipage}[t]{3.5cm}
\begin{math}
x^2 - 4x = 12 \\
x^2 - 4x + 4 = 16 \\
(x - 2)^2 = 4^2 \\
x-2 = 4 \vee x-2 = -4 \\
x = 6 \vee x = -2
\end{math}
\end{minipage}\hspace*{4mm}
\begin{minipage}[t]{4cm}
\begin{math}
x^2 - 4x = 12 \\
x^2 - 4x - 12 = 0 \\
\hspace*{3mm}\begin{array}{|@{\:}r@{\:}c@{\:}l}
D &=& (-4)^2 - 4 \cdot 1 \cdot - 12 \\
  &=& 64 \\
\sqrt{D} &=& \sqrt{64} = 8 \\
\end{array} \\
x = \frac{4 + 8}{2} \vee x = \frac{4 - 8}{2} \\
x = 6 \vee x = -2
\end{math}
\end{minipage}
\end{center}
\caption{Three possible derivations for a quadratic equation}
\label{figure:derivations}
\end{figure}

Depending on the context, a \emph{teacher} may want to control the way in which 
a particular (set of) exercise(s) is solved. For example, a certain 
exercise should be 
solved without using the quadratic formula, or without the technique of 
completing the square 
(because it may not be part of the course material). Controlling the 
solution space not only has an effect on the diagnosis of an intermediate term 
entered by a learner, it also influences the generation of hints and worked-out 
solutions. A strategy that combines multiple solution techniques will often not 
be of help, since hints and worked-out solutions might refer to techniques 
unknown to the learner, or techniques that should not be used.

\subsection{Case study: changing the level of detail}
\label{case2}

While doing an exercise, a \emph{learner} wants to increase the level of detail
that is presented by the learning environment, 
i.e., the granularity of the steps. For example, the learner might find 
the step in which $x=\frac{1}{2}\sqrt{32}$ is simplified to $x=2\sqrt{2}$ hard 
to understand, even though familiarity with simplifying roots is assumed. 
According to the principle of glass-box computation the learner should be able 
to inspect the calculations within this step. An extreme scenario in the other 
direction is a learner who is only interested in the final answer, not in the 
intermediate answers.

\subsection{Case study: changing the number system}
\label{case3}

A \emph{teacher} wants to allow complex numbers in solutions for polynomial 
equations, instead of real numbers. In the setting with real numbers, a 
negative discriminant (or a squared term that has to be negative) leads to no 
solutions. According to the principle of cognitive fidelity, the software should 
solve the problem with complex numbers or with real numbers, depending on the 
teacher's preference. However, the approach to solve an equation, that is, the 
rewrite strategy, is not changed significantly. Therefore, reuse of the existing 
strategy is desirable. A similar scenario would be to restrict the numbers in an 
equation to rationals only, without introducing square roots.

\subsection{Case study: creating new exercises from existing parts}
\label{case4}

Rewrite strategies can often be extended to deal with a new class of exercises 
by performing some steps beforehand or afterwards. In the case of solving an 
equation with a polynomial of degree 3 or higher, one could try to reduce the 
problem to a quadratic equation. This equation can then be handled by an 
existing strategy for solving quadratic equations. Ideally, such a composite 
strategy is already defined and available. If not, a mathematical \emph{learning 
environment} (or a \emph{teacher} using it) should be able to assemble the 
strategy from existing parts, and use it in the same way as a predefined 
strategy.

%JJ: we have to shorten the paper a bit? I wouldn't know what to refer to here.
%In an earlier project~\cite{intelligentfeedback-surf} we already observed that 
%many linear algebra exercises involve reducing a matrix, with some pre-processing 
%(or post-processing) steps. For such a family of related exercises, it is useful 
%to allow rewrite strategies and rules to be assembled outside the scope of the 
%domain reasoner.

Another scenario is a collection of rules that has to be applied exhaustively to
solve an exercise. Although exhaustive application of rules results in a very
simple rewrite strategy, many interesting problems can be solved in this way. It
should therefore be possible for a \emph{teacher} using the learning environment
to take or specify such a collection, and to construct a strategy out of it.

\subsection{Case study: customizing an exercise with a student model}
\label{case5}

Advanced \emph{learning environments}, such as {\activemath}, maintain a student
model containing information about the skills and capabilities of the learner.
Such a student model can be used for different purposes, including task
selection and reporting the progress of a learner. Because the model contains
detailed knowledge about the level of the learner, it is desirable to use
this knowledge and to customize the domain reasoner accordingly. For example, a
learner that understands Gaussian elimination can perform this method as a
single step when determining the inverse of a matrix. On the contrary, 
beginners in linear algebra should see the intermediate steps.

Obviously, diagnoses from the domain reasoners should also be used to update the
student model. In both cases, the domain reasoner and the learning environment
need a shared understanding of the knowledge items, such as the rewrite rules
and the rewrite strategies. The exchange of information in both directions
suggests that the two parts should be tightly integrated.

\section{Concepts and representation of knowledge}
\label{section:concepts}

This section discusses the three concepts that are the foundation of our 
approach: rewrite rules, rewrite strategies, and views for defining canonical 
forms. These concepts not only assist in reasoning about exercises at a 
conceptual level, they are also the core abstractions in the implementation of 
the domain reasoners. We give a brief introduction to each of the concepts, and 
point out how they represent knowledge appearing in mathematical textbooks. 
Furthermore, we highlight the properties of the concepts. In the last part of 
this section we discuss how the concepts come together in defining an exercise.

\subsection{Rewrite rules}

Rewrite rules specify how terms can be manipulated in a sound way, and are often
given explicitly in textbooks. Well-known examples are rewriting 
$AB = 0$ into $A = 0 \mathbin{\:\vee\:} B = 0$, the quadratic formula, 
and associativity of
addition. These rules constitute the steps in worked-out solutions. Soundness of
rules can be checked with respect to some semantic interpretation of a formula.
Such an interpretation can be context-specific (e.g., $x^2 \!=\! -3$ gives no 
solutions for \ensuremath{\Varid{x}} in~$\mathbb{R}$).

Rewrite rules are atomic actions that can be implemented in code. Clearly, this
gives the implementer of the rule the full power of the underlying programming
language. An alternative is to specify rules with a left-hand side and a
right-hand side, and to rely on unification and substitution of terms to do the
transformation~\cite{vannoort:rewriting}. This is common practice in 
term rewrite systems
(TRS)~\cite{baader-TRS}. 
We allow rewrite rules to yield multiple results.

\subsection{Rewrite strategies}

Simple problems can be solved by applying a set of rules exhaustively 
(for instance, when the set of rules is confluent), but this is generally not 
the case.
A rewrite strategy~\cite{rewritestrategiesforexercises} 
guides the process of applying rewrite rules to solve a particular class of 
problems. Recipes for solving a certain type of problem can be found in 
textbooks, but they are often not precise enough for the purpose of building a 
domain reasoner. Given a collection of worked-out solutions by an expert, one 
can try to infer the strategy that was used 
(although typically only one possible derivation is covered).
 
Rewrite strategies are built from rewrite rules, with combinators for sequences 
and choices (\ensuremath{\mathrel{{<}\hspace{-0.35em}*\hspace{-0.35em}{>}}} and \ensuremath{\mathrel{{<}\hspace{-0.05em}|\hspace{-0.05em}{>}}}, respectively). The fixed point combinator \ensuremath{\Varid{fix}} 
allows for repeating parts. Labels can be placed at arbitrary places in the 
strategy, marking substrategies of interest. From a strategy description, 
multiple derivations may be generated or recognized. 
%A detailed description of rewrite strategies can be found in earlier 
%papers~\cite{strategiesforexercises,rewritestrategiesforexercises}.

Since strategies only structure the order in which rewrite rules are applied, 
soundness of a derivation follows directly from the soundness of the rules 
involved. Note that a strategy not only prescribes which rule to apply, but also 
where (that is, to which subterm). Also, strategies are designed with a specific 
goal in mind. A strategy for quadratic equations, for instance, is expected to 
rewrite an equation until the variable is isolated. The solved form that a 
strategy is supposed to reach is the strategy's post-condition. Likewise, a 
strategy may have certain assumptions about the starting term (e.g., the 
equation must be quadratic, or only a single variable is involved), which is its 
pre-condition.

\subsection{Views and canonical forms}

Canonical forms and notational conventions are an integral part of courses on
mathematics. Examples of conventions in writing down a polynomial are the order
of its terms (sorted by the degree of the term), and writing the coefficient in
front of the variable. Such conventions also play a role when discussing
equations of the form $ax^2 + bx = 0$: it is likely that $-3x + x^2 = 0$ is
considered an instance of the form, although the expression $1x^2 + (-3)x$ is
rather atypical. These implicit assumptions make that standard rewriting
techniques do not apply directly.

Canonical forms and notational conventions can be captured in a
view~\cite{canonicalforms}, which consists of a partial function for matching,
and a (complete) function for building. Matching may result in a value of a
different type, such as the pair \ensuremath{(\mathbin{-}\mathrm{3},\mathrm{5})} for the expression \ensuremath{\mathbin{-}(\mathrm{3}\mathbin{-}\mathrm{5})}. In this
example, the interpretation of the pair would be addition of both parts. Having
a value of a different type after matching can be useful when specifying a
rewrite rule: the pair \ensuremath{(\mathbin{-}\mathrm{3},\mathrm{5})}, for instance, witnesses that an addition was
recognized at top-level. Building after matching gives the canonical form, and
this composed operation must therefore be idempotent. A view is assumed to
preserve a term's semantics.

Primitive views can be composed into compound views, in two different ways.
Firstly, views are closely related to the arrow interface~\cite{PatersonRA:fop},
and its bidirectional variant. The combinators of this interface can be used
for combining views, such as using views in succession. Secondly, views can be
parameterized with another view. Consider a view for expressions of the form $ax
+ b$, returning a pair of expressions for $a$ and $b$. Another view can then be
used for these two parts (e.g., a view for rational numbers). Essentially, this
pattern of usage corresponds to having higher-order views.
%JJ: too detailed?
%\footnote{A similar view can be constructed with the arrow combinators for the
%given example. However, parameterized views (or higher-order views) correspond
%more closely to the intuition of having a normal-form within a normal-form.}.
Views can be used in different ways:
\begin{itize}
   \item as a rewrite rule, reducing a term to its canonical form (if possible); 
   \item as a predicate, checking whether a term has a canonical form;
   \item as an equivalence relation, comparing the canonical forms of two terms.
\end{itize}

\subsection{Exercises}

The three fundamental concepts for constructing domain reasoners discussed in
this section are all we need to support a general set of feedback services
(Section~\ref{services}). Instances of the concepts are grouped together in an
\emph{exercise} containing all the domain-specific (and exercise-specific)
functionality.

The most prominent component of an exercise is its rewrite strategy. In addition 
to the rewrite rules that are present in the strategy, more rules can be added 
to the exercise for the purpose of being recognized, including buggy rules for 
anticipating common mistakes. Predicates are needed for checking whether a term 
is a suitable starting term that can be solved by the strategy, and whether a 
term is in solved form. These two predicates can be defined as views. For 
diagnosing intermediate answers, we need an equivalence relation to compare a 
submission with a preceding term. This relation can be specified as a view. 
Besides checking student submissions, this view can be used as an 
internal consistency 
check, validating the soundness of the rewrite rules. One more view is needed 
that checks whether two terms are similar enough to be considered the same. This 
view is used to bring intermediate terms produced by a strategy to their 
canonical forms.

What remains to be supplied for an exercise is its metadata, such as an 
identifier that can serve as a reference, and a short description. 
For certain domains it is convenient to have a dedicated 
parser and pretty-printer for the terms involved. For external tools, however, 
interchanging abstract syntax (as opposed to concrete syntax), such as OpenMath 
objects~\cite{openmath} for mathematical domains, is the preferred way of 
communication, avoiding the need for a parser and pretty-printer. 
Although not of primary importance, it can be convenient to have a 
randomized term generator for the exercise.

\section{Adaptation and configuration}
\label{section:config}

This section discusses how users can adapt and customize the exercises that are 
offered by a domain reasoner. A user has to be able to 
inspect the internals of the components of an exercise, to adapt and replace 
these components, and to create new exercises. 
We briefly discuss the consequences of 
applying the glassbox principle to our components. We then propose 
representations for rewrite rules, rewrite strategies, and views. These 
representations are an essential part of the communication with a domain 
reasoner. Strategy configurations are introduced for conveniently adapting 
existing strategies. We conclude by returning to our case studies, and show how 
they can be addressed.

\subsection{The glassbox principle}

The glassbox principle expresses that you should be able to see all steps 
leading to a final answer. This is possible with our current 
services, but you cannot query the specifics of a rule that was applied, or 
examine the structure of the rewrite strategy. From the perspective of a 
learning environment, rewrite strategies and rules are still black boxes 
delivering some result. Ideally, the components involved are transparent as 
well, and adhere to the glassbox principle.

Exposing the internals of a component has the advantage that more details become
available for the learning environment, and for other external tools. These
details can be communicated to learners, or to teachers writing feedback
messages. The information can also be used for documentation, visualization of
rewrite strategies, analyses, and much more. Once a domain reasoner 
supports a representation, it can be extended to interpret 
descriptions that are passed to it.
As a result, exercises can be adapted in new, unforeseen ways.

However, there is a trade-off in making components fully transparent. The need
for a representation that can be communicated restricts the way components can
be specified. The developer of a domain reasoner can no longer take advantage of
the facilities offered by the underlying programming language, which may
negatively affect performance, for example. For our own domain reasoners, we are
gradually working towards transparency.

\subsection{Representing rewrite rules}
\label{section:repr-rules}

Consider the rewrite rule 
$AB = AC \rightarrow A=0 \vee B=C$.
In this rule, \ensuremath{\Conid{A}}, \ensuremath{\Conid{B}}, and \ensuremath{\Conid{C}} are meta-variables 
representing arbitrary expressions. A rule that is written in this way can be 
seen as a Formal Mathematical Property (FMP), a concept introduced by the 
OpenMath standard~\cite{openmath} to specify properties 
of symbols that are defined in content 
dictionaries. The OpenMath standard also supports explicit quantification of 
meta-variables by means of the \text{\tt forall} binder in the \text{\tt quant1} dictionary. We 
can thus use FMPs to represent the rewrite rules of our domain 
reasoners\footnote{Instead of using FMPs, we could have introduced our own 
representation, in which case we would still need quantification, meta-variables, 
and a pair of terms.}. 
Likewise, buggy rules can be communicated as FMPs, except that the 
meta-variables are existentially quantified.
Indeed, many of our rewrite rules can also be found in a content dictionary 
as an FMP. 

Unfortunately, not all rules can be represented with a left and right-hand 
side straightforwardly. 
Keep in mind that the representation of a rule should closely correspond to  
how it is perceived by a learner. We give some examples that challenge
this approach.
\begin{itize}
   \item Some steps correspond to primitive operations, such as replacing 
      \ensuremath{\mathrm{3}\mathbin{+}\mathrm{5}} by \ensuremath{\mathrm{8}}, or reducing $\frac{10}{15}$ to $\frac{2}{3}$. Special
      support is needed for these operations.
   \item Rewrite rules should not have meta-variables on the 
      right-hand side that do not appear on the left~\cite{baader-TRS}.
      Conceptually, however, such rules do exist as an intermediate step, such 
      as the rule for scaling a fraction ($\frac{A}{B} \rightarrow \frac{AC}
      {BC}$), as a preparatory step for adding it to another fraction. This 
      rule also shows that rules can have side conditions ($C \not= 0$),
      which can be expressed in an FMP.
   \item Generalizations of rules involving a variable number of terms require 
      special support. An example of such a rule is
      $A(B_1 + \ldots + B_n) \rightarrow AB_1 + \ldots + AB_n$.
   \item In an earlier paper~\cite{canonicalforms} we have argued that rules are 
      specified in 
      the context of a view, yet there is no support for views in the rewrite 
      rules. 
\end{itize}
These cases can only be circumvented partially by having explicit 
support for views in rewrite rules (i.e., associate a new symbol with a view,
and specialize the unification procedure for that symbol), or by using 
strategies as a representation for rules (recall that rules can return 
multiple results).

With this representation for rewrite rules, learning environments can 
communicate new rules to the domain reasoner, thereby extending it. Essentially, 
this turns the domain reasoner into a rewrite rule \emph{interpreter}. When 
allowing dynamic extension of a domain, it may no longer be possible to 
guarantee (or check) the soundness of rules. Also, care should be taken that 
the new rules do not result in excessive computations.

\subsection{Representing rewrite strategies}
\newcommand{\xml}{XML}

Rewrite strategies are specified using a small set of combinators, such as 
\ensuremath{\mathrel{{<}\hspace{-0.35em}*\hspace{-0.35em}{>}}} for sequence, and \ensuremath{\mathrel{{<}\hspace{-0.05em}|\hspace{-0.05em}{>}}} for choice. Additional combinators are 
defined in terms of this small set (e.g., \ensuremath{\Varid{repeat}}), resulting in a combinator 
library with common patterns. For example, consider the strategy specification 
for solving a linear equation, in which both sides of the equation 
are first rewritten into their basic form 
$ax + b$ (the preparation step).
% lineq, prepare, basic :: LabeledStrategy (Equation Expr)
\begin{hscode}\SaveRestoreHook
\column{B}{@{}>{\hspre}l<{\hspost}@{}}%
\column{10}{@{}>{\hspre}c<{\hspost}@{}}%
\column{10E}{@{}l@{}}%
\column{13}{@{}>{\hspre}l<{\hspost}@{}}%
\column{14}{@{}>{\hspre}l<{\hspost}@{}}%
\column{E}{@{}>{\hspre}l<{\hspost}@{}}%
\>[B]{}\Varid{lineq}{}\<[10]%
\>[10]{}\mathrel{=}{}\<[10E]%
\>[13]{}\Varid{label}\;\text{\tt \char34 linear~equation\char34}\;(\Varid{prepare}\mathrel{{<}\hspace{-0.35em}*\hspace{-0.35em}{>}}\Varid{basic}){}\<[E]%
\\
\>[B]{}\Varid{prepare}{}\<[10]%
\>[10]{}\mathrel{=}{}\<[10E]%
\>[13]{}\Varid{label}\;\text{\tt \char34 prepare~equation\char34}\;{}\<[E]%
\\
\>[13]{}\hsindent{1}{}\<[14]%
\>[14]{}(\Varid{repeat}\;(\Varid{merge}\mathrel{{<}\hspace{-0.05em}|\hspace{-0.05em}{>}}\Varid{distribute}\mathrel{{<}\hspace{-0.05em}|\hspace{-0.05em}{>}}\Varid{removeDivision})){}\<[E]%
\\
\>[B]{}\Varid{basic}{}\<[10]%
\>[10]{}\mathrel{=}{}\<[10E]%
\>[13]{}\Varid{label}\;\text{\tt \char34 basic~equation\char34}\;{}\<[E]%
\\
\>[13]{}\hsindent{1}{}\<[14]%
\>[14]{}(\Varid{try}\;\Varid{varToLeft}\mathrel{{<}\hspace{-0.35em}*\hspace{-0.35em}{>}}\Varid{try}\;\Varid{conToRight}\mathrel{{<}\hspace{-0.35em}*\hspace{-0.35em}{>}}\Varid{try}\;\Varid{scaleToOne}){}\<[E]%
\ColumnHook
\end{hscode}\resethooks
This strategy specification is declarative and compositional, which allows for 
an almost literal translation into an {\xml} equivalent. The {\xml} fragment 
for the \ensuremath{\Varid{lineq}} strategy is given below:
\begin{small}
\begin{tabbing}\tt
~~~~\char60{}label~name\char61{}\char34{}linear~equation\char34{}\char62{}\\
\tt ~~~~~~\char60{}sequence\char62{}\\
\tt ~~~~~~~~\char60{}label~name\char61{}\char34{}prepare~equation\char34{}\char62{}\\
\tt ~~~~~~~~~~\char60{}repeat\char62{}\char60{}choice\char62{}\\
\tt ~~~~~~~~~~~~\char60{}rule~name\char61{}\char34{}merge\char34{}\char47{}\char62{}\\
\tt ~~~~~~~~~~~~\char60{}rule~name\char61{}\char34{}distribute\char34{}\char47{}\char62{}\\
\tt ~~~~~~~~~~~~\char60{}rule~name\char61{}\char34{}remove~division\char34{}\char47{}\char62{}\\
\tt ~~~~~~~~~~\char60{}\char47{}choice\char62{}\char60{}\char47{}repeat\char62{}\\
\tt ~~~~~~~~\char60{}\char47{}label\char62{}\\
\tt ~~~~~~~~\char60{}label~name\char61{}\char34{}basic~equation\char34{}\char62{}~\char46{}\char46{}\char46{}~\char60{}\char47{}label\char62{}\\
\tt ~~~~~~\char60{}\char47{}sequence\char62{}\\
\tt ~~~~\char60{}\char47{}label\char62{}
\end{tabbing}
\end{small}
An {\xml} tag is introduced for each combinator, and labels and rules have
attributes for storing additional information. The strategy combinators for sequence 
and choice are associative, and therefore we let their corresponding tags
have arbitrary many children, instead of imposing a nested structure. The declarative
nature of rewrite strategies makes that such a convention does not interfere 
with the meaning of the strategy, i.e., it is easy to reason about strategies.

An important design decision in the representation of rewrite strategies is
which of the derived combinators to support in \xml, and which not. For
instance, \ensuremath{\Varid{repeat}\;\Varid{s}} is defined as \ensuremath{\Varid{many}\;\Varid{s}\mathrel{{<}\hspace{-0.35em}*\hspace{-0.35em}{>}}not\;\Varid{s}}, where \ensuremath{\Varid{many}} and \ensuremath{not}
are also derived combinators. Instead of introducing the tag \text{\tt \char60{}repeat\char62{}}, we
could use \ensuremath{\Varid{repeat}}'s definition, giving a \text{\tt \char60{}sequence\char62{}} tag at top-level. Fewer
tags make it easier for other tools to process a strategy description. On the
other hand, tools can take advantage of the extra tags (e.g., a tool for
visualizing strategies). Hence, we decide to support most of the 
combinators in our library.
% Alternatively, tools have to infer often occurring patterns themselves.

Rules are referenced by name in a strategy. Similarly, known (sub)strategies can
be included as well. This is particularly helpful for assembling new strategies
from existing parts (both rules and strategies). Under the assumption that the
parts have already been defined, we can give a concise strategy description
for the running example:
\begin{small}
\begin{tabbing}\tt
~~~~\char60{}label~name\char61{}\char34{}linear~equation\char34{}\char62{}\\
\tt ~~~~~~\char60{}sequence\char62{}\\
\tt ~~~~~~~~\char60{}strategy~name\char61{}\char34{}prepare~equation\char34{}\char47{}\char62{}\\
\tt ~~~~~~~~\char60{}strategy~name\char61{}\char34{}basic~equation\char34{}\char47{}\char62{}\\
\tt ~~~~~~\char60{}\char47{}sequence\char62{}\\
\tt ~~~~\char60{}\char47{}label\char62{}
\end{tabbing}
\end{small}

The {\xml} representation paves the way for learning environments to offer their
own rewrite strategies, turning the domain reasoner into an interpreter for 
strategy descriptions. Interpreting strategies raises issues concerning 
the correctness of the strategy (the post-condition it should establish), 
and in particular termination when rewriting with the strategy. Experience
has shown that specifying rich strategies is a difficult and error-prone
activity, for which offline analysis and testing capabilities are very helpful.

\subsection{Configuring rewrite strategies}

New rewrite strategies can be defined from scratch, but often a small change to 
an existing strategy suffices. Strategy \emph{configurations} offer an 
alternative (and simpler) way to adapt strategies. With such a 
configuration, a sequence of transformations can be applied to a strategy.

A useful transformation is to \emph{remove} a specific part of a strategy, such
that it is not used in a derivation. This can be carried out by replacing the 
part (substrategy or rule) by \ensuremath{\Varid{fail}}, which is the unit element of the choice
combinator. When you remove a rule,
you risk that an exercise can no longer be solved. 
The inverse transformation is to 
\emph{reinsert} a part that was marked as removed.

Another transformation is based on the fact that strategies are special 
instances of rewrite rules, since they can be performed in a single step.
Thus, strategies can be \emph{collapsed} into a rule, contributing to just 
one step in a derivation. The inverse operation is to \emph{expand} a rewrite
rule and turn it into a strategy.
% , for which it has to be known how the expansion is done.

The \ensuremath{\Varid{hide}} transformation makes a rule implicit, or the rules in a 
rewrite strategy. An implicit rule behaves normally, except that it does not 
show up as a step in a derivation. Implicit rules can be used to perform certain 
simplification steps automatically, and are comparable to so-called 
administrative rules~\cite{rewritestrategiesforexercises}. The inverse of \ensuremath{\Varid{hide}}
is the \ensuremath{\Varid{reveal}} transformation.

The properties \ensuremath{\Varid{removed}}, \ensuremath{\Varid{collapsed}}, and \ensuremath{\Varid{hidden}} correspond to the 
transformations described above, and they can be assigned to subexpressions 
in a strategy description. 
The properties are translated to attributes in an {\xml} representation.
The three inverse transformations appear as attributes set to false, which
is their default value. The following \xml\ snippet illustrates this approach:
\begin{small}
\begin{tabbing}\tt
~~~~\char60{}label~name\char61{}\char34{}basic~equation\char34{}~collapsed\char61{}\char34{}true\char34{}\char62{}~\char46{}\char46{}\char46{}~\char60{}\char47{}label\char62{}
\end{tabbing}
\end{small}
Note that this still requires the whole strategy to be communicated,
including the part that is collapsed. To
circumvent this, we introduce an {\xml} tag for each transformation with a
target specifying where the transformation should be applied. The following
{\xml} fragment takes the original strategy for solving a linear equation, and
applies the \emph{collapse} transformation to the substrategy labeled \ensuremath{\text{\tt \char34 basic~equation\char34}}:
\begin{small}
\begin{tabbing}\tt
~~~~\char60{}collapse~target\char61{}\char34{}basic~equation\char34{}\char62{}\\
\tt ~~~~~~~\char60{}strategy~name\char61{}\char34{}linear~equation\char34{}\char47{}\char62{}\\
\tt ~~~~\char60{}\char47{}collapse\char62{}
\end{tabbing}
\end{small}
Transformations can be combined, and if nested, the innermost is applied 
first. Because strategy transformations are pure functions, they can be 
freely mixed with the ``regular'' strategy combinators.

Instead of removing a part, we have seen cases where the opposite was requested by 
a teacher: a certain rule (or substrategy) must be used. This can be done by 
selectively removing parts, and making sure that the mandatory part is used in all 
cases. For convenience, we offer a \emph{mustuse} transformation doing exactly 
that, which can be used as the other transformations. A weaker variant is to 
express a preference for using a rule: this boils down to replacing some choice 
strategy combinators by the left-biased choice combinator (written~\ensuremath{\mathrel{\triangleright}}, 
see~\cite{rewritestrategiesforexercises}). The \emph{prefer} transformation 
guarantees that 
the same set of exercises can be solved by the strategy, which is not the 
case for \emph{mustuse}. The final transformation we discuss is to 
\emph{replace} a part of the strategy by something else. This transformation 
takes a target to be replaced, the replacement (first child), and the strategy 
in which the replacement has to take place (second child).

\subsection{Representing views}
\label{section:repr-views}

Finding a representation for a view is arguably more difficult than finding 
one for a rewrite rule or strategy. Since a view is just a pair of 
functions, it is unclear how its internal structure could be represented 
in general, other than its implementation in the underlying programming
language. We discuss two special cases: a view defined as a confluent set
of rewrite rules, and a view specified as a rewrite strategy.
Compound views are represented by introducing an explicit representation
for the arrow combinators (as was done for the strategy combinators), 
and a representation for the application
of higher-order views.

Some views can be defined as a confluent set of rewrite rules, in particular  
views for simplifying the complete term, and not just the top-level nodes
of the term. 
The view's function for 
matching applies the set of rules: its function for building is simply the 
identity function. Such a view can be represented by listing the rules.
Note that confluence ensures that the view returns a canonical form.

Views can also be specified by a rewrite strategy for the view's match function 
and its build function. This is more sophisticated than providing a confluent 
set of rewrite rules, because the strategy can control in a precise way how the 
rules should be applied. The strategy language has a fixed point combinator for 
expressing general recursion. This makes it plausible that many views can be 
written as a strategy. The operations of a view must be idempotent,
and this property must be checked for views that are represented by a rewrite
strategy.

\subsection{Case studies revisited}
\label{section:revisit}

We briefly revisit the five case studies. The 
teacher in case study~\ref{case1} wants to control how an exercise is solved, 
for example by disallowing certain rules or techniques. A strategy 
configuration provides this functionality by means of the \emph{remove}, 
\emph{mustuse}, and \emph{prefer} transformations. 
The second case (a learner customizing the 
level of details) is handled likewise: parts in the strategy that have been 
collapsed can be expanded, or the other way around. To see yet more detail, 
implicit rules can be made explicit, or rules can be replaced by a rewrite 
strategy that is doing the same. For example, the quadratic formula introduces
a square root, which is simplified immediately because it 
is not the focus of the exercise. Normalizing 
expressions involving roots is, however, a topic on its own, for which a 
rewrite strategy is available. We can plug-in this strategy to increase the 
level of detail in solving an equation with the quadratic formula.

Changing the underlying number system 
in an exercise (case study~\ref{case3}) is not trivial. 
Consider using complex numbers for solving a 
quadratic equation. To start with, some support for the basic operations 
on complex numbers is needed (e.g., addition and multiplication).
This can best be captured in a view. Ideally, a view for complex numbers is 
already present in the domain reasoner. If not, the view can 
be specified as a rewrite strategy. This view can be used for bringing an 
expression with complex numbers to its canonical form.
Furthermore, additional 
rewrite rules are added to the exercise, such as $i^2 \rightarrow -1$. 
% $(a + bi)(c + di) \rightarrow (ac - bd) + (bc + ad)i$. 
These new rules can be inserted in the strategy for quadratic equations, whereas 
other rules are excluded (e.g., the rule that a square root of a negative 
number leads to no solution).
The subtle part of 
this case study is that the views used in the strategy's rewrite rules may also 
have to change, in particular if they involve calculations with numbers. 
Composing higher-order views (e.g., a view for polynomials parameterized 
over the type of its coefficients) alleviates this issue.

Case study~\ref{case4} is solved by interpreting rewrite strategies that are 
assembled by the learning environment. Not all rewrite rules are representable, 
which currently limits what can be done without changing the domain reasoner. 
The last case study involves customizing the level of 
detail in an exercise, which is highly desirable for adaptive learning systems. 
Based on the student model, a strategy configuration must be generated by 
the learning environment. 

\section{Conclusions, related and future work}
\label{section:conclusions}

We have shown why adapting domain reasoners is very desirable in the context of 
mathematical learning environments. By explicitly representing the fundamental
concepts used in domain reasoners, we can let users adapt and configure a class
of exercises in a domain reasoner. We use OpenMath to represent mathematical
expressions and rewrite rules, but we have designed our own \xml\ language for
specifying rewrite strategies, and transformations on these strategies. Our
strategy language is very similar to the tactic languages used in theorem
proving~\cite{bundyproofplans,aspinall2010}, and has the same expressive power.

Several authors discuss adaptation of various aspects in learning
environments~\cite{pahl03,DBLP:conf/ectel/UllrichLM09}, but we are not aware of
previous work on configuring and adapting domain reasoners. Hierarchical
proofs~\cite{cheikhrouhou00,aspinall2010}, which represent proofs at different
levels of abstraction, are related to turning a strategy into a rule and vice
versa. As far as we found, hierarchical proofs are not used to recognize proving
steps made by a student.

We have indicated some challenges in representing rewrite rules and views 
(sections~\ref{section:repr-rules} and~\ref{section:repr-views}), and these 
cases require further investigation. Even though we are striving for domain 
reasoners that are fully transparent (i.e., that have an explicit 
representation), we think that hybrid solutions, in which only certain parts can 
be adapted, are a conceivable compromise. We plan to investigate how the 
facilities for adapting domain reasoners can best be offered to a teacher or a 
domain expert, and what skills are reasonable to expect from such a user.

\ifthenelse{\boolean{arxiv}}{}{\vspace*{-1.5mm}}%
\paragraph{Acknowledgements.} 
This work was made possible by the \MathBridge\ project of the Community
programme eContent\emph{plus}, and by the SURF Foundation
(\url{http://www.surf.nl}), the higher education and research partnership
organization for Information and Communications Technology (ICT), for the NKBW2
project. The paper does not represent the opinion of the Community, and the
Community is not responsible for any use that might be made of information
contained in this paper. We acknowledge discussions with Peter Boon, George
Goguadze, and Wim van Velthoven. We thank Alex Gerdes and the anonymous
reviewers for commenting on an earlier version of this paper.

%needed for the page limit
\ifthenelse{\boolean{arxiv}}{}{\vspace*{-2mm}\begin{small}}
\bibliographystyle{plain}
\bibliography{AdaptingDomainReasoners}

\providecommand{\noopsort}[1]{}
\begin{thebibliography}{10}

\bibitem{getalenruimte}
C.~Admiraal~et al.
\newblock {\em Getal \& Ruimte}.
\newblock EPN, Houten, The Netherlands, 2009.

\bibitem{aspinall2010}
D.~Aspinall, E.~Denney, and C.~L\"{u}th.
\newblock Tactics for hierarchical proof.
\newblock {\em Mathematics in Computer Science}, 3(3):309--330, 2010.

\bibitem{baader-TRS}
F.~Baader and T.~Nipkow.
\newblock {\em Term Rewriting and All That}.
\newblock Cambridge University Press, 1997.

\bibitem{mathxpert}
M.J. Beeson.
\newblock Design principles of {Math{P}ert}: Software to support education in
  algebra and calculus.
\newblock In N.~Kajler, editor, {\em Computer-Human Interaction in Symbolic
  Computation}, pages 89--115. Springer, 1998.

\bibitem{1576039}
R.~Bradford, J.H. Davenport, and C.J. Sangwin.
\newblock A comparison of equality in computer algebra and correctness in
  mathematical pedagogy.
\newblock In {\em MKM'09}, pages 75--89. Springer, 2009.

\bibitem{bundyproofplans}
A.~Bundy.
\newblock The use of explicit plans to guide inductive proofs.
\newblock In {\em CADE}, pages 111--120, 1988.

\bibitem{cheikhrouhou00}
L.~Cheikhrouhou and V.~Sorge.
\newblock {PDS} -- a three-dimensional data structure for proof plans.
\newblock In {\em ACIDCA}, 2000.

\bibitem{mathdox}
A.~Cohen, H.~Cuypers, E.~Reinaldo~Barreiro, and H.~Sterk.
\newblock Interactive mathematical documents on the web.
\newblock In {\em Algebra, Geometry and Software Systems}, pages 289--306.
  Springer, 2003.

\bibitem{doormanetal2009}
M.~Doorman, P.~Drijvers, P.~Boon, S.~van Gisbergen, and K.~Gravemeijer.
\newblock Design and implementation of a computer supported learning
  environment for mathematics.
\newblock In {\em Earli 2009 SIG20 invited Symposium Issues in designing and
  implementing computer supported inquiry learning environments}, 2009.

\bibitem{gerdes-08}
A.~Gerdes, B.~Heeren, J.~Jeuring, and S.~Stuurman.
\newblock Feedback services for exercise assistants.
\newblock In D.~Remenyi, editor, {\em ECEL}, pages 402--410. Acad. Publ. Ltd.,
  2008.

\bibitem{DBLP:conf/mkm/Goguadze09}
G.~Goguadze.
\newblock Representation for interactive exercises.
\newblock In {\em MKM'09}, volume 5625 of {\em LNCS}, pages 294--309, 2009.

\bibitem{canonicalforms}
B.~Heeren and J.~Jeuring.
\newblock Canonical forms in interactive exercise assistants.
\newblock In {\em MKM'09}, volume 5625 of {\em LNCS}, pages 325--340. Springer,
  2009.

\bibitem{rewritestrategiesforexercises}
B.~Heeren, J.~Jeuring, and A.~Gerdes.
\newblock Specifying rewrite strategies for interactive exercises.
\newblock {\em Mathematics in Computer Science}, 3(3):349--370, 2010.

\bibitem{DBLP:conf/icaisc/MelisS04}
E.~Melis and J.~Siekmann.
\newblock Activemath: An intelligent tutoring system for mathematics.
\newblock In {\em ICAISC}, volume 3070 of {\em LNCS}, pages 91--101. Springer,
  2004.

\bibitem{vannoort:rewriting}
T.~{\noopsort{Noort}}van~Noort, A.~Rodriguez~Yakushev, S.~Holdermans,
  J.~Jeuring, B.~Heeren, and J.P. {Magalh\~aes}.
\newblock A lightweight approach to datatype-generic rewriting.
\newblock {\em Journal of Functional Programming}, 2010.
\newblock To appear.

\bibitem{pahl03}
C.~Pahl.
\newblock Managing evolution and change in web-based teaching and learning
  environments.
\newblock {\em Computers \& Education}, 40(2):99--114, 2003.

\bibitem{PatersonRA:fop}
R.~Paterson.
\newblock Arrows and computation.
\newblock In J.~Gibbons and O.~de~Moor, editors, {\em The Fun of Programming},
  pages 201--222. Palgrave, 2003.

\bibitem{openmath}
The~OpenMath Society.
\newblock {The OpenMath Standard}.
\newblock \url{http://www.openmath.org/standard/index.html}, 2006.

\bibitem{DBLP:conf/ectel/UllrichLM09}
C.~Ullrich, T.~Lu, and E.~Melis.
\newblock A new framework for dynamic adaptations and actions.
\newblock In {\em EC-TEL}, volume 5794 of {\em LNCS}, pages 67--72, 2009.

\end{thebibliography}
\ifthenelse{\boolean{arxiv}}{}{\end{small}}

\end{document}